\documentclass[twocolumn]{aastex631}

\shorttitle{Spectroscopic orbits of subsystems VIII.}

\begin{document}

\renewcommand{\topfraction}{1.0}
\renewcommand{\bottomfraction}{1.0}
\renewcommand{\textfraction}{0.0}

\newcommand{\kms}{km~s$^{-1}$\,}
\newcommand{\msun}{$M_\odot$\,}

\title{Spectroscopic orbits of subsystems in  multiple
  stars. VIII.}

\author{Andrei Tokovinin}
\affiliation{Cerro Tololo Inter-American Observatory | NSF's NOIRLab
Casilla 603, La Serena, Chile}
\email{andrei.tokovinin@noirlab.edu}

\begin{abstract}
Periods,  eccentricities, and masses  in hierarchical  stellar systems
inform us  on the formation  and early evolution of  these fascinating
objects.  To  complement  the  multiplicity statistics  of nearby  solar-type
stars,  19  new  spectroscopic   orbits  of  inner  subsystems  in  15
hierarchies  (10 triples  and 5  quadruples) are  determined  based on
high-resolution echelle spectra collected during several years.  While
previous papers  of this series contained  mostly short-period orbits,
here most periods  are on the order  of a year. The  main components of
these  hierarchies are  HIP 7852,  9148, 12548,  21079,  24320, 27970,
34212,  56282,  57860, 76400,  76816,  81394,  96284,  100420, and  HD
108938. Noteworthy systems are HIP 12548 and 24230 (hierarchies of 2+2
architecture with low-mass spectroscopic secondaries), HIP 56282 (a
planetary-type 3+1 hierarchy), and HIP 27970 (a compact triple with
periods of 15 and 1049 days). 
\end{abstract}



\section{Introduction}
\label{sec:intro}

Observations of  spectroscopic subsystems  in nearby  solar-type stars
are  motivated by  the  desire  to determine  their  periods and  mass
ratios,   complementing  statistics   of  hierachies   in  the   solar
neighborhood  \citep{FG67b}.   Many  subsystems  discovered  by,  e.g.
\citet{N04}, lack  orbits and  therefore confuse the  statistics. This
paper presents results of the long-term program conducted at the 1.5 m
telescope  at  Cerro Tololo  with  the  CHIRON high-resolution  optical
echelle   spectrograph.    Previous   publications  of   this   series
\citep{chiron1,chiron2,chiron2a,chiron3,chiron4,chiron5,chiron5a,chiron6,chiron7}
focused   primarily  on   the   short-period  subsystems.    Continued
accumulation of the spectra allows  calculation of orbits with periods
on the order of a year presented here.  Knowledge of these orbits adds
missing parameters  of the nearby  hierarchies and contributes  to the
collection   of    such   data   in   the    Multiple   Star   Catalog
\citep[MSC,][]{MSC}.

\begin{figure}
\epsscale{1.2}
\plotone{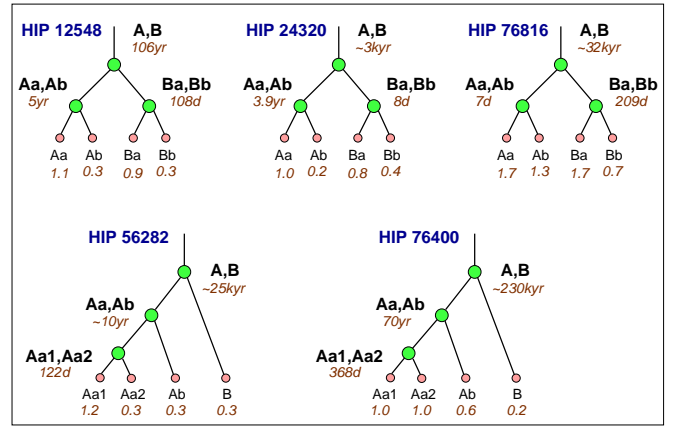}
\caption{Architecture of five quadruple systems. Green circles denote
  subsystems, small pink circles -- individual stars with masses
  indicated below.  
\label{fig:mobile}
}
\end{figure}

\begin{deluxetable*}{c c rr   l cc rr r c }
\tabletypesize{\scriptsize}     
\tablecaption{Basic parameters of observed multiple systems
\label{tab:objects} }  
\tablewidth{0pt}                                   
\tablehead{                                                                     
\colhead{WDS} & 
\colhead{Comp.} &
\colhead{HIP} & 
\colhead{HD} & 
\colhead{Spectral} & 
\colhead{$V$} & 
\colhead{$V-K_s$} & 
\colhead{$\mu^*_\alpha$} & 
\colhead{$\mu_\delta$} & 
\colhead{RV} & 
\colhead{$\varpi$\tablenotemark{a}} \\
\colhead{(J2000)} & 
 & &   &  
\colhead{type} & 
\colhead{(mag)} &
\colhead{(mag)} &
\multicolumn{2}{c}{ (mas yr$^{-1}$)} &
\colhead{(km s$^{-1}$)} &
\colhead{(mas)} 
}
\startdata
01410$-$0524    & A  & 7852  & 10353 & G5V     & 8.48  & 1.69 & $-$69 &$-$27 & 0.7 & 19.44: \\ 
                & B  & \ldots & \ldots & K5    & 11.84 & 3.46 & $-$69 &$-$26 & $-$0.4 & 22.03 \\ 
01579$-$2851    & A  & 9148  & 12068   & G3V   & 8.27  & 1.67 & 212   & 57   & 23.2   & 12.45: \\ 
                & B  & \ldots &\ldots& M:      & 13.07 & 3.18 & 212   & 57   & 23.5   & 11.24 \\ 
02415$-$7128    & AB & 12548 & 17215 & G6V     & 7.82  & 1.76 & 158 & $-$61 & 45.1 & 18.6:\tablenotemark{b} \\
04311$-$4522    & AB & 21079 & 28904 & G3V     & 8.29  & 1.71 & $-$50 & 17  & $-$3.5 & 21.87: \\ 
05131$-$5959    & A  & 24320 & 34377 & G5V     & 8.87  & 1.73 &  30  & 126  & 16.1 & 15.87:   \\ 
                & B  &\ldots & \ldots &\ldots & 10.13  & 2.55 & 19   & 125  & 16.5 & 16.43:   \\   
05550$-$1256    & A  & 27970 & 39899 & F7V    & 7.73   & 1.28 & $-$23 & $-$42 & $-$12.0 & 15.78: \\ 
07056$-$7116    & A  & 34212 & 55197 & F8V    & 7.66   & 1.30 &$-$5   & 267   & 52.1   & 14.39: \\ 
                & B  & \ldots& \ldots& \ldots & 16.91  & 5.24 &$-$5   & 270   &\ldots & 16.55 \\ 
11323$-$0025    & A  & 56282 & 100269& F8V    & 8.09   & 1.35 & 43   & $-$161 & 8.3   & 15.13: \\ 
                & B  & \ldots& \ldots& \ldots& 16.23   & 5.15 & 41   &$-$162  & \ldots & 16.28 \\ 
11520$-$4357    & A & 57860 & 103076& G5V    & 9.02   & 1.64 & $-$20 &$-$21  & 38.5    & 12.09 \\ 
                & B & \ldots & \ldots & \ldots  & 9.27   & \ldots & $-$24 &$-$25  & 36.4    & 12.31 \\
12314$-$5659    &AB & \ldots & 108938 & G8V   & 8.98   & 2.17 &  45   & 7     & 39.0    & 6.35 \\ 
15362$-$0623    & A & 76400  & 139059 & G6V   & 7.90   & 1.62 & $-$103&$-$179 & $-$7.3  & 15.28: \\
                & B & \ldots & \ldots & \ldots& 17.82  & 5.81 &$-$99  &$-$174 & \ldots  & 15.29 \\
15410$-$1449    & A & 76816  & 139864 & F8V   & 9.47   & 1.62 & $-$26 & $-$1  & $-$39.2 & 3.25 \\ 
                & B & \ldots & \ldots & \ldots& 9.74   & 2.50 &$-$25  &$-$2   & $-$40.3 & 3.32 \\ 
16374$-$6133    & A & 81394  & 149261 & F5    & 8.54   & 1.28 &  11   & $-$17 & $-$15.3 & 5.02 \\ 
                & B & 81395  & 149262 & F2    & 9.22   & 1.17 &  11   & $-$17 & $-$14.1 & 4.96 \\ 
19583$-$5154    & A & 98294  & 188557 & A9IV  & 7.40   & 0.54 &  46   &$-$44  & $-$5.0  & 10.41 \\ 
                & B & 98278  & 188534 & F6V   & 8.17   & 1.18 &  46   &$-$43  &$-$6.3   & 10.49 \\
20218$-$3654    & A & 100420 & \ldots & F6V   & 8.31   & 0.94 &   2   &$-$16  & 49.1    & 5.61: \\   
                & B & \ldots & 193465 & \ldots& 9.60   & 1.29 &   2   &$-$18  & 49.1    & 5.65 
\enddata
\tablenotetext{a}{Proper motions and parallaxes are 
  from Gaia EDR3 \citep{gaia3}. Colons mark parallaxes biased by subsystems, in which case the PMs are
 from \citet{Brandt2018}.  }
\tablenotetext{b}{ Hipparcos parallax \citep{HIP2}.}
\end{deluxetable*}

The hierarchical     systems     studied     here    are     listed     in
Table~\ref{tab:objects}. The  data are collected from  Simbad and Gaia
\citep[][hereafter EDR3]{gaia3},   the  radial   velocities   (RVs)  are   mostly
determined in this work. The  first column gives the Washington Double
Star \citep[WDS,][]{WDS} code based on the J2000 coordinates.  The HIP
and HD  identifiers, spectral types, photometric  and astrometric data
refer either to the individual stars or to the unresolved subsystems. 
Parallaxes potentially biased by unresolved subsystems are marked by
colons. 

This paper is organized similarly to  the previous ones.  The data and
methods  are  briefly  outlined in  section~\ref{sec:obs},  where  the
orbital  elements  are  also  given.  Then  each  multiple  system  is
discussed    in     section~\ref{sec:obj}.     Figure~\ref{fig:mobile}
illustrates the architecture of quadruple systems (triples are obvious
and do  not need illustration).   The paper closes with  discussion of
the results in section~\ref{sec:disc}.

\section{Observations and Data Analysis}
\label{sec:obs}

\subsection{Spectroscopic Observations}

The spectra used here were taken with the 1.5 m telescope sited at the
Cerro Tololo  Inter-American Observatory (CTIO) in  Chile and operated
by  the   Small  and   Medium  Aperture  Telescopes   Research  System
\href{http://www.astro.yale.edu/smarts/}{(SMARTS)}         Consortium.
Fifteen hours  of observing  time were allocated  to this  program per
semester.  Observations  were made  with the fiber-fed  CHIRON optical
echelle  spectrograph  \citep{CHIRON,Paredes2021}   by  the  telescope
operators in  service mode.  The  spectra taken with the  image slicer
have a resolution of 85,000.  They  are reduced by the standard CHIRON
pipeline.  The  wavelength calibration  is based on  the thorium-argon
lamp spectra taken after each object.

The RVs  are determined  from Gaussian  fits to  the cross-correlation
function (CCF)  of echelle orders  with the  binary mask based  on the
solar spectrum,  as detailed in  \citet{chiron1}.  The RVs  derived by
this method rely  on the instrument wavelength  calibration and should
be on the  absolute scale, so no instrumental  corrections are applied
\citep[an  offset of  $+0.15$ km~s$^{-1}$  relative to  established RV
  standards was found in][]{chiron3}.  The RV errors depend on several
factors such  as the width  and contrast of  the CCF dip,  presence of
other dips,  and signal-to-noise  ratio.  The  rms residuals  from the
orbits can be as  low as 0.01 \kms, but typically  are between 0.1 and
0.5 \kms for the systems studied  here.  I assign the RV errors (hence
weights)  to  match roughly  the  residuals,  with larger  errors  for
blended or  noisy dips.  Some blended  CCFs are  fitted by  fixing the
width  or amplitude  of  individual components  determined from  other
spectra with  better-separated dips. Otherwise, a  heavily blended dip
is  fitted by  a  single  Gaussian, and  the  resulting  biased RV  is
assigned a large error and a  low weight in the orbit fit.

The width of the CCF dip is related to the projected rotation velocity
$V \sin i$, while its  area depends on the spectral type, metallicity,
and, for  blended spectra  of several stars,  on the  relative fluxes.
Table~\ref{tab:dip}  lists average parameters  of the  Gaussian curves
fitted to the CCF dips.   It gives the number of averaged measurements
$N$ (blended CCFs are ignored), the dip amplitude $a$, its dispersion
$\sigma$, the product  $a \sigma$ proportional to the  dip area (hence
to the relative flux), and the projected rotation velocity $V \sin i$,
estimated  from   $\sigma$  by   the  approximate  formula   given  in
\citep{chiron1}  and valid  for $\sigma  < 12$  \kms.  The  last column
indicates  the presence  or absence  of the  lithium 6708\,\AA  ~line in
individual components.

\begin{deluxetable*}{l l c cccc c}    
\tabletypesize{\scriptsize}     
\tablecaption{CCF parameters
\label{tab:dip}          }
\tablewidth{0pt}                                   
\tablehead{                                                                     
\colhead{HIP/HD} & 
\colhead{Comp.} & 
\colhead{$N$} & 
\colhead{$a$} & 
\colhead{$\sigma$} & 
\colhead{$a \sigma$} & 
\colhead{$V \sin i$ } & 
\colhead{Li}
\\
 &  &  & &
\colhead{(km~s$^{-1}$)} &
\colhead{(km~s$^{-1}$)} &
\colhead{(km~s$^{-1}$)} &
\colhead{  6708\AA}
}
\startdata
7852      & Aa &  7 & 0.359 &  4.228 & 1.52   & 4.5  & Y \\
9148      & Aa &  7 & 0.370 &  3.634 & 1.35   & 2.3  & Y \\
9148      & Ab &  7 & 0.090 &  3.764 & 0.34   & 2.9  & N? \\
12548     & Aa & 15 & 0.330 &  3.773 & 1.25   & 2.9  & N \\
12548     & Ba & 15 & 0.130 &  3.677 & 0.48   & 2.5  & N \\
21079     & Aa &  8 & 0.319 &  4.549 & 1.45   & 5.4  & Y \\ 
24320     & Aa & 14 & 0.401 &  4.198 & 1.68   & 4.4  & Y? \\
24320     & Ba & 13 & 0.414 &  4.354 & 1.80   & 4.9  & N \\
27970     & Aa1& 6  & 0.299 &  3.751 & 1.12   & 2.8  & Y! \\ 
34212     & Aa & 12 & 0.287 &  4.490 & 1.29   & 5.3  & N \\
56282     & Aa1& 15 & 0.297 &  3.933 & 1.17   & 3.6  & Y!  \\  
57860     & Aa & 7  & 0.187 &  3.551 & 0.66   & 1.8  & N \\
57860     & Ab & 7  & 0.141 &  3.539 & 0.50   & 1.8  & N \\
57860     & B  & 7  & 0.134 &  3.596 & 0.48   & 2.1  & N \\ 
108938    & Aa & 2  & 0.324 &  3.490 & 1.13   & 1.4   & N \\
108938    & Ab & 2  & 0.048 &  3.130 & 0.15   & 0     & N \\
108938    & B  & 2  & 0.061 &  3.845 & 0.23   & 3.2   & N \\ 
76400     & Aa & 9  & 0.234 &  3.735 & 0.88   & 2.8   & N \\
76400     & Ab & 9  & 0.233 &  3.726 & 0.87   & 2.7   & N \\
76816     & Ba & 13 & 0.507 &  3.806 & 1.93   & 3.1   & Y? \\
81394     & Ba & 13 & 0.086 &  7.471 & 0.65   & 12.0  & N \\ 
81394     & Bb & 13 & 0.098 &  6.064 & 0.59   & 9.0   & N \\ 
98278     & Ba & 8  & 0.070 & 10.68  & 0.75   & 18.2  & Y \\
98278     & Bb & 8  & 0.029 & 12.75  & 0.37   & 22.1: & \ldots \\
100420    & Aa & 12 & 0.215 &  4.628 & 1.00   & 5.7   & N \\
100420    & Ab & 12 & 0.070 &  4.218 & 0.30   & 4.5   & N \\
100420    & B & 3  & 0.241 &   5.435 & 1.31   & 7.6   & \ldots  
\enddata 
\end{deluxetable*}

\subsection{Orbit Calculation}

The  orbital  elements   and  their  errors  are   determined  by  the
least-squares fits with weights  inversely proportional to the adopted
RV    errors.    The    IDL   code    {\tt   ORBIT}\footnote{Codebase:
  \url{http://www.ctio.noirlab.edu/\~atokovin/orbit/}              and
  \url{https://doi.org/10.5281/zenodo.61119}      }      was      used
\citep{orbit}. In  some triple  systems, the orbits  of the  outer and
inner subsystems are  fitted jointly to the RVs  and, where available,
position  measurements using  a  modification of  the  same code  {\tt
  ORBIT3} \citep{ORBIT3} described by \citet{TL2017}. Both codes allow
to fix some orbital elements  to avoid degeneracies (e.g. for circular
or  face-on  orbits)  or  to  cope with  insufficient  data  (e.g.  an
incomplete coverage of the outer orbit).

\begin{deluxetable*}{l l cccc ccc c c}    
\tabletypesize{\scriptsize}     
\tablecaption{Spectroscopic orbits
\label{tab:sborb}          }
\tablewidth{0pt}                                   
\tablehead{                                                                     
\colhead{HIP/HD} & 
\colhead{System} & 
\colhead{$P$} & 
\colhead{$T$} & 
\colhead{$e$} & 
\colhead{$\omega_{\rm A}$ } & 
\colhead{$K_1$} & 
\colhead{$K_2$} & 
\colhead{$\gamma$} & 
\colhead{rms$_{1,2}$} &
\colhead{$M_{1,2} \sin^3 i$} 
\\
& & \colhead{(d)} &
\colhead{(JD +24,00,000)} & &
\colhead{(deg)} & 
\colhead{(km~s$^{-1}$)} &
\colhead{(km~s$^{-1}$)} &
\colhead{(km~s$^{-1}$)} &
\colhead{(km~s$^{-1}$)} &
\colhead{ (${\cal M}_\odot$) } 
}
\startdata
7852 & Aa,Ab &  1177.9 & 58615.0 & 0.349 & 260.5 & 7.320 & \ldots & 0.677  & 0.014  & 0.98: \\
     &    & $\pm$3.4 & $\pm$16.0 & $\pm$0.015 & $\pm$3.6 & $\pm$0.116 & \ldots & $\pm$0.105  & \ldots & 0.43 \\
9148 & Aa,Ab & 1272.3 & 58531.6 & 0.0 & 0.0 & 7.711 & 9.550 & 23.163  & 0.16  & 0.37  \\
     &    & $\pm$6.0 & $\pm$2.2 & fixed & fixed & $\pm$0.099 & $\pm$0.099 & $\pm$0.057  & 0.16  & 0.30 \\
12548 & Aa,Ab & 1851.9    & 59023.7  & 0.544      & 151.7     & 5.02      & \ldots & 45.1 & 0.11 & 1.07: \\
     &       & $\pm$18.4 & $\pm$29.6 & $\pm$0.040 & $\pm$5.4 & $\pm$0.66 & \ldots & fixed & \ldots & 0.32 \\
12548 & Ba,Bb & 108.18 & 59241.648  & 0.823      & 201.3     & 21.36     & \ldots & 47.60    & 0.26 & 0.89: \\
     &       & $\pm$0.10 & $\pm$0.087 & $\pm$0.005 & $\pm$0.7 & $\pm$0.28 & \ldots & $\pm$0.12 & \ldots & 0.31  \\
21079&  Aa,Ab & 217.04   & 58044.39  & 0.491      & 336.6    & 5.949      & \ldots & $-$3.525   & 0.024  & 0.98: \\ 
     &       & $\pm$0.23 & $\pm$1.66 & $\pm$0.008 & $\pm$1.1  & $\pm$0.103     & \ldots & $\pm$0.041     & \ldots & 0.16  \\
24320 & Aa,Ab & 1430.3   & 58800.2   & 0.393     & 136.6     & 3.46      & \ldots    & 16.14       & 0.034 & 1.00: \\
  &       & $\pm$10.6 & $\pm$10.2    & $\pm$0.018 & $\pm$1.5  & $\pm$0.03 & \ldots & $\pm$0.05    & \ldots & 0.19  \\
24320 & Ba,Bb & 7.94292 & 58190.82   & 0.0240    & 287.5      & 36.76     & \ldots & 16.46        & 0.038 & 0.82: \\
     &     & $\pm$0.00002 & $\pm$0.10 & $\pm$0.0005 & $\pm$4.5 & $\pm$0.11     & \ldots & $\pm$0.02 & \ldots & 0.39  \\
27970 & Aa1,Aa2 & 15.3230  & 56600.996  & 0.504      & 197.8    & 25.48       & \ldots  & \ldots      & 0.22 & 1.25: \\
     &    & $\pm$0.0001 & $\pm$0.041 & $\pm$0.007  & $\pm$1.3  & $\pm$0.36   & \ldots & \ldots     & \ldots & 0.35 \\
27970 & Aa,Ab & 1049.4 & 57207.6    & 0.359       & 231.4     & 6.13       & \ldots  & $-$11.95   & 0.22 & 1.60: \\
  &       & $\pm$4.3 & $\pm$11.6    & $\pm$0.028    & $\pm$5.5  & $\pm$0.20   & \ldots & $\pm$0.11   & \ldots & 0.44 \\
34212 & Aa,Ab & 1246.7 & 57902.5 & 0.163 & 192.3 & 7.265 & \ldots & 52.051  & 0.015 &  1.24: \\
      &   & $\pm$2.9 & $\pm$12.7 & $\pm$0.005 & $\pm$3.6 & $\pm$0.058 & \ldots & $\pm$0.031 & \ldots  & 0.53 \\
56282 & Aa1,Aa2 & 122.177 & 58945.35 & 0.046 & 264.9 & 8.284          & \ldots & \ldots  & 0.066 &  1.14: \\
      & & $\pm$0.058 & $\pm$2.23     & $\pm$0.006 & $\pm$6.6 & $\pm$0.063 & \ldots & \ldots & \ldots  & 0.24  \\
56282 & Aa,Ab  & 3650 & 57233 & 0.056 & 228.5 & 2.527          & \ldots & 8.254  & 0.066 &  1.38: \\
       &     & fixed & $\pm$1271  & $\pm$0.066 & $\pm$123.4 & $\pm$0.520 & \ldots & $\pm$0.455 & \ldots  & 0.25 \\
57860    & Aa,Ab & 891.9 & 58137.7  & 0.101        & 6.0       & 11.75      & 12.42 & 38.47  & 0.11  & 0.66 \\
  &       & $\pm$2.7   & $\pm$ 9.3  & $\pm$0.005   & $\pm$4.0  & $\pm$0.06  & $\pm$0.09 & $\pm$0.04  & 0.06 & 0.62  \\
108938    & Aa,Ab  & 342.48     & 59321.34   & 0.517        & 208.6     & 17.72      & 21.66     & 39.01      & 0.22 & 0.75 \\
  &       & $\pm$1.20 & $\pm$0.82   & $\pm$0.018  & $\pm$1.3   & $\pm$0.32  & $\pm$0.32 & $\pm$0.30  & 0.41 & 0.61 \\
76400     & Aa,Ab  & 368.51  & 58235.66  & 0.137       & 192.0      & 18.64      & 18.70     & $-$7.28    & 0.05 & 0.96 \\
 &       & $\pm$0.74   & $\pm$3.66    & $\pm$0.039  & $\pm$3.7   & $\pm$0.47  &$\pm$0.47 & $\pm$0.71   & 0.04 & 0.96 \\
76816 & Ba,Bb  &  208.918 & 59185.16 & 0.420 & 43.2 & 15.265 & \ldots & $-$40.323    &  0.016   & 1.67: \\
      &   & $\pm$0.011 & $\pm$0.34 & $\pm$0.005 & $\pm$0.9 & $\pm$0.082 & \ldots & $\pm$0.062 & \ldots & 0.68 \\
81395 &  Ba,Bb & 224.804 & 58319.14 & 0.418 & 147.6 & 11.14 & 12.04 & $-$14.13     &  0.34 & 0.11 \\
      &   & $\pm$0.028 & $\pm$0.46 & $\pm$0.006 & $\pm$1.2 & $\pm$0.21 & $\pm$0.08 & $\pm$0.06  & 0.12  & 0.10 \\
98278    & Ba,Bb & 236.15     & 58754.64     & 0.574        & 245.2      & 27.22      & 27.95   & $-$6.32    & 1.20 & 1.14 \\
   &       & $\pm$0.33 & $\pm$0.92    & $\pm$0.008          & $\pm$1.0   & $\pm$0.42  & $\pm$0.32 & $\pm$0.24  & 0.57 & 1.11 \\
100420   & Aa,Ab  & 790.6    & 58395.68     & 0.372       & 73.0       & 14.78      & 17.52  & 49.08         & 0.11 & 1.20 \\      
     &       & $\pm$0.46   & $\pm$0.88     & $\pm$0.003    & $\pm$0.5  & $\pm$0.06 & $\pm$0.06 & $\pm$0.03   & 0.13 & 1.00 
\enddata 
\end{deluxetable*}


\begin{deluxetable}{r l c rrr c }    
\tabletypesize{\scriptsize}     
\tablecaption{Radial velocities and residuals (fragment)
\label{tab:rv}          }
\tablewidth{0pt}                                   
\tablehead{                                                                     
\colhead{HIP} & 
\colhead{System} & 
\colhead{Date} & 
\colhead{RV} & 
\colhead{$\sigma$} & 
\colhead{(O$-$C) } &
\colhead{Comp.}  \\
\colhead{HD} & & 
\colhead{(JD +24,00,000)} &
\multicolumn{3}{c}{(km s$^{-1}$)} &
\colhead{Instr.}
}
\startdata
  7852 &Aa,Ab  &  53695.0000 &   -6.46 &    0.20 &   -0.02 &  aC \\
  7852 &Aa,Ab  &  54046.6120 &    7.18 &    0.10 &   -0.01 &  aH \\
  7852 &Aa,Ab  &  57319.6410 &   -6.93 &    0.20 &    0.01 &  a \\
  7852 &Aa,Ab  &  57983.8210 &    2.30 &    0.20 &    0.00 &  a \\
  7852 &Aa,Ab  &  58745.7420 &    6.96 &    0.20 &    0.03 &  a
\enddata 
\tablenotetext{}{(This table is available in its entirety in
  machine-readable form). Instrument codes:  
C -- \citet{Chubak2012};
D -- \citet{Desidera2006}; 
H -- HARPS;
L -- Las Campanas \citep{LCO}; 
R -- Radial Velocity Meter \citep{Gorynya2018}.
}
\end{deluxetable}

Table~\ref{tab:sborb} gives  elements of  the spectroscopic  orbits in
standard notation.  Its  last column contains the masses  $M \sin^3 i$
for double-lined binaries.  For single-lined  systems, the mass of the
primary star (listed  with colons) is estimated from  its absolute $V$
magnitude, and the  minimum mass of the secondary  that corresponds to
the    90\degr   ~inclination    is    derived    from   the    orbit.
Table~\ref{tab:rv},   published  in   full  electronically,   provides
individual RVs.   The Hipparcos or HD  number of the primary  star and
the system  identifier (components joined  by comma) in the  first two
columns define  the pair.  Then  follow the  Julian date, the  RV, its
adopted error $\sigma$ (blended CCF  dips are assigned larger errors),
and the residual  to the orbit (O$-$C).  The last  column specifies to
which  component this  RV refers  ('a' for  the primary,  'b' for  the
secondary)  and gives  references to  the published  RVs.  The  RVs of
other   visual  components   are   provided,   for  completeness,   in
Table~\ref{tab:rvconst}.  It  contains the  HIP number,  the component
letter, the  Julian date, and the  RV.  The less accurate  RVs derived
from blended dips are marked by colons.

\begin{deluxetable}{r l r r }    
\tabletypesize{\scriptsize}     
\tablecaption{Radial velocities of other components
\label{tab:rvconst}          }
\tablewidth{0pt}                                   
\tablehead{                                                                     
\colhead{HIP/HD} & 
\colhead{Comp.} & 
\colhead{Date} & 
\colhead{RV}   \\ 
 & & 
\colhead{(JD +24,00,000)} &
\colhead {(km s$^{-1}$)}  
}
\startdata
57860 & B &   58193.7198 & 36.034 \\
57860 & B &   58194.6254 & 36.005 \\
57860 & B &   58195.6408 & 35.973 \\
57860 & B &   58510.8194 & 35.994: \\
57860 & B &   58539.7575 & 35.961: \\ 
57860 & B &   58559.6517 & 35.942: \\ 
57860 & B &   58648.6288 & 36.129: \\ 
57860 & B &   59403.4624 & 36.000 \\
108938 & B &  54576.6567 & 38.212  \\
108938 & B &  59316.6752 & 37.679  \\
108938 & B &  59323.6711 & 37.707  \\
108938 & B &  59404.5562 & 38.227:  \\
100420 & B &  57985.5413 & 49.437 \\  
100420 & B &  57986.5850 & 49.489 \\  
100420 & B &  58549.9038 & 49.408  
\enddata 
\end{deluxetable}

\subsection{Other Data}

I use  here astrometry and photometry  from the Gaia third  early data
release,  EDR3  \citep{gaia3}.   For   multiple  systems,  it  can  be
compromised by  acceleration and/or unresolved companions  because the
current  astrometric   reductions  assume  single  stars.    The  RUWE
parameter  (Reduced   Unit  Weight   Error)  captures   the  excessive
astrometric noise, helping to  identify suspicious astrometry in EDR3.
Most (but not all) stars with  subsystems studied here have RUWE $>2$.
Uncertain    Gaia    parallaxes    are    marked    by    colons    in
Table~\ref{tab:objects}.  Detection  of astrometric subsystems  by the
difference $\Delta \mu$ (so-called  PMA) between the short-term proper
motion (PM)  measured by  Gaia and the  long-term PM  $\mu_{\rm mean}$
deduced from  the Gaia  and Hipparcos positions  by \citet{Brandt2018}
continued the  work by \citet{MK05} at  a new level of  accuracy.  For
stars with large RUWE, I use the long-term PMs determined by Brandt in
place of the PMs measured by Gaia.

For   some   systems,   spectroscopy   is  complemented   by   speckle
interferometry of the outer pairs. Most speckle observations used
here were made at the Southern Astrophysical Research Telescope (SOAR)
and are  referred to  in the  text simply as  `SOAR data'.  The latest
observations  and  references  to  older  publications  are  found  in
\citet{Tokovinin2021}. Apart  from the position  measurements, speckle
interferometry  provides  differential   photometry  of  close  visual
pairs. For  binaries wider than $\sim$1\arcsec, it  is complemented by
the  Gaia photometry. Effective temperatures estimated by Gaia
give the  spectral types  of the visual  components, checked  by their
$V-K$  colors. Spectral  types  and absolute  magnitudes  are used  to
estimate masses  of the  stars using standard  main-sequence relations
from \citet{Pecaut2013}.

\section{Individual Objects}
\label{sec:obj}

For double  and triple-lined systems,  Figures in this section  show a
typical CCF (the  Julian date and individual components  are marked on
the  plot) together  with the  RV curve  representing the  orbit.  For
single-lined  binaries, only  the RV  curves are  plotted.  In  the RV
curves,  green squares  denote the  primary component,  blue triangles
denote the secondary  component, while the full and  dashed lines plot
the orbit. Typical error bars are smaller than the symbols.  Masses of
stars are estimated from absolute  magnitudes, orbital periods of wide
pairs       ---      from       their      projected       separations
\citep[see][]{MSC}. Semimajor axes of the spectroscopic subsystems are
computed using the third Kepler's  law, and the photocenter amplitudes
are evaluated based on the estimated masses and fluxes.

\subsection{HIP 7852 (Triple)}

\begin{figure}[ht]
\plotone{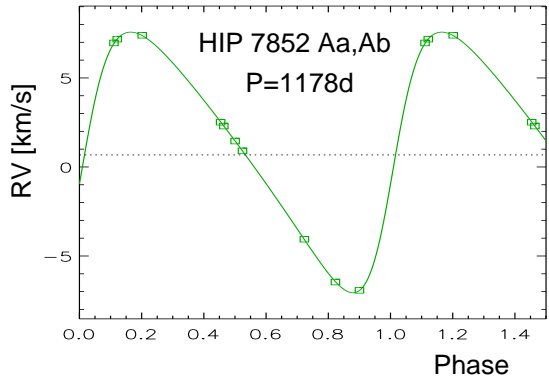}
\caption{RV curve of HIP 7852 Aa,Ab.
\label{fig:7852} 
}
\end{figure}

This is a chromospherically  active star EY~Cet \citep{BoroSaikia2018}
located  at 45\,pc  distance.   \citet{Luck2018} used  the archive  of
high-resolution spectra taken with the HAPRS spectrometer to determine
the effective temperature of 5641\,K (spectral type G5V), a mild metal
deficiency [Fe/H]=$-$0.13, and  the projected rotation of $V  \sin i =
5.8$   \kms,   in   good   agreement   with   4.5   \kms   listed   in
Table~\ref{tab:dip}. The HARPS  and CHIRON spectra show  a presence of
lithium despite the slow rotation and the age of 5.8\,Gyr estimated by
Luck.  The common proper motion  (CPM) companion B ($V=11.84$ mag, K5,
UCAC2~29946080)  at 125\arcsec  ~separation is  listed in  the WDS  as
TOK~228.  Stars A and B are located on the main sequence.

The  astrometric acceleration  of  star A  was  detected by  Hipparcos
\citep{MK05} and by \citet{Brandt2018}; it explains the huge RUWE=15.6
in Gaia EDR3.   The star has  been  placed  on the  CHIRON  program  in 2015  to
determine the period of Aa,Ab.  Eight single-lined spectra accumulated
to date  indicate $P=3.2$\,yr.  This orbit  (Figure~\ref{fig:7852}) is
strengthened by including the RV deduced from the HARPS spectrum taken
in 2006 (JD 2454046.612) and recovered from the ESO archive and the RV
measured  in  2005  (JD  2453695.0) by  \citet{Chubak2012}.   The  rms
residuals are only  0.014 \kms. The RV  from Gaia  is  not used here
because its mean epoch  is not known and the error  is large (1.9 \kms)
owing to variability.

Star Aa has an estimated mass of 0.98 \msun, the minimum mass of Ab is
0.43 \msun,  and the semimajor  axis of the  inner orbit is  47.5 mas.
The  minimum   Ab  mass  corresponds   to  the  photocenter   axis  of
14.5\,mas. An astrometric orbit will determined by Gaia, leading to an
accurate mass of Ab.

\subsection{HIP 9148 (Triple)}

\begin{figure}
\epsscale{1.0}
\plotone{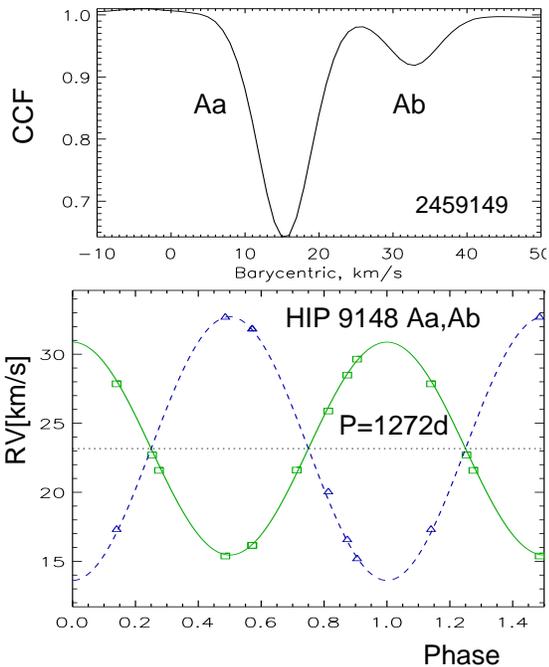}
\caption{CCF (top)  and RV curve (bottom) of HIP 9148 Aa,Ab. 
\label{fig:9148}
}
\end{figure}

The outer  pair A,B, at  41\farcs5 separation, has been  identified as
CPM by  \citet{Luyten1941} and is listed  in the WDS as  LDS~61.  Gaia
confirms that this  high-PM pair is bound. Star B  ($V=13.07$ mag) has
an  estimated mass  of 0.55  \msun.  Its  EDR3 parallax  of 11.24  mas
(distance 89\,pc) is not biased by  the subsystem. In contrast, star A
is a double-lined binary \citep{N04}. It has a large $\Delta \mu$, the
EDR3 RUWE is also large, 17.6,  and the EDR3 parallax of 12.45\,mas is
biased.   Like  the previous  object,  the  star is  chromospherically
active \citep{BoroSaikia2018} and has a detectable lithium line, although
the projected axial rotation is slow.

Ten CHIRON spectra acquired since 2017 show double CCFs with unequal
dips (Figure~\ref{fig:9148}). The  magnitude difference between Aa and
Ab deduced from the dip areas  is 1.5 mag, hence the $V$ magnitudes of
Aa  and Ab are  8.51 and  10.01 mag,  respectively, and  correspond to
main-sequence  stars  with  masses  of  1.23  and  0.95  \msun  (ratio
0.77). Aa is brighter than inferred  from its spectral type G3V, it is
located slightly above the  main sequence (evolved), while  star B is just
below.

The orbit of Aa,Ab  with a period of 3.5 yr is  circular.  A free fit
gives $e=0.017  \pm 0.020$ and does  not reduce the  residuals, so the
circular orbit  is enforced. The spectroscopic  mass ratio $q=0.81$
is similar to the ratio  of photometricaly estimated masses, which are
much larger  than the spectroscopic  masses $M \sin^3 i$.   This means
that  the  Aa,Ab  orbit  has  a large  inclination  of  42\degr.   The
semimajor  axis of  Aa,Ab is  33\,mas, and  the  estimated photocenter
amplitude is about 8 mas.

\subsection{HIP 12548 (Quadruple)}

The  star HIP  12548  (HD 17215,  G6V,  $V=7.82$ mag)  belongs to  the
immediate  solar  neighborhood:  the  revised  Hipparcos  parallax  of
18.6$\pm$0.8 mas puts it at the 54\,pc distance. The Gaia DR2 parallax
of 15.8$\pm$0.8 mas  is inaccurate, and EDR3 gives  no parallax.  This
is a  visual binary B~1923  (WDS J02415$-$7128) known since  1931; its
orbit with a  period of $\sim$100 yr and a  semimajor axis of 0\farcs6
is now  almost fully covered and well constrained (see below).

\citet{N04} found  that the  spectrum is double-lined  and, therefore,
there is a spectroscopic subsystem.  They even determined a mass ratio
of 0.26  by assuming that the  lines belong to the  same spectroscopic
system.   However,   their  paper  does  not   contain  individual  RV
measurements, making  it impossible  to use  or check  these findings.
Only single lines were seen in 2008 at Las Campanas \citep{LCO} and in
2017 with CHIRON. Occasionally, a double-lined spectrum is observed in
a visual binary near periastron  of its eccentric orbit; the existence
of a spectroscopic subsystem remained thus questionable.

The visual binary  was monitored by speckle interferometry  at SOAR in
2008-2021,  allowing us to  improve its  orbit.  To  check for  potential
spectroscopic subsystems, the object was  placed on the CHIRON program
in  2017. The  observed  CCF  profiles varied  from  narrow single  to
asymmetric  or double,  confirming  the existence  of a  spectroscopic
subsystem.   Furthermore,  it  became  clear  that  the  RVs  of  both
components seen in the spectra are variable.  Both visual components A
and  B  are  close  binaries  with  periods of  5  yr  and  108  days,
respectively,  and  this  is  a quadruple  system  of  2+2  hierarchy.
Discovery of the true nature of  this seemingly common visual pair was
possible only with high-resolution spectroscopy.

\begin{figure}[ht]
\plotone{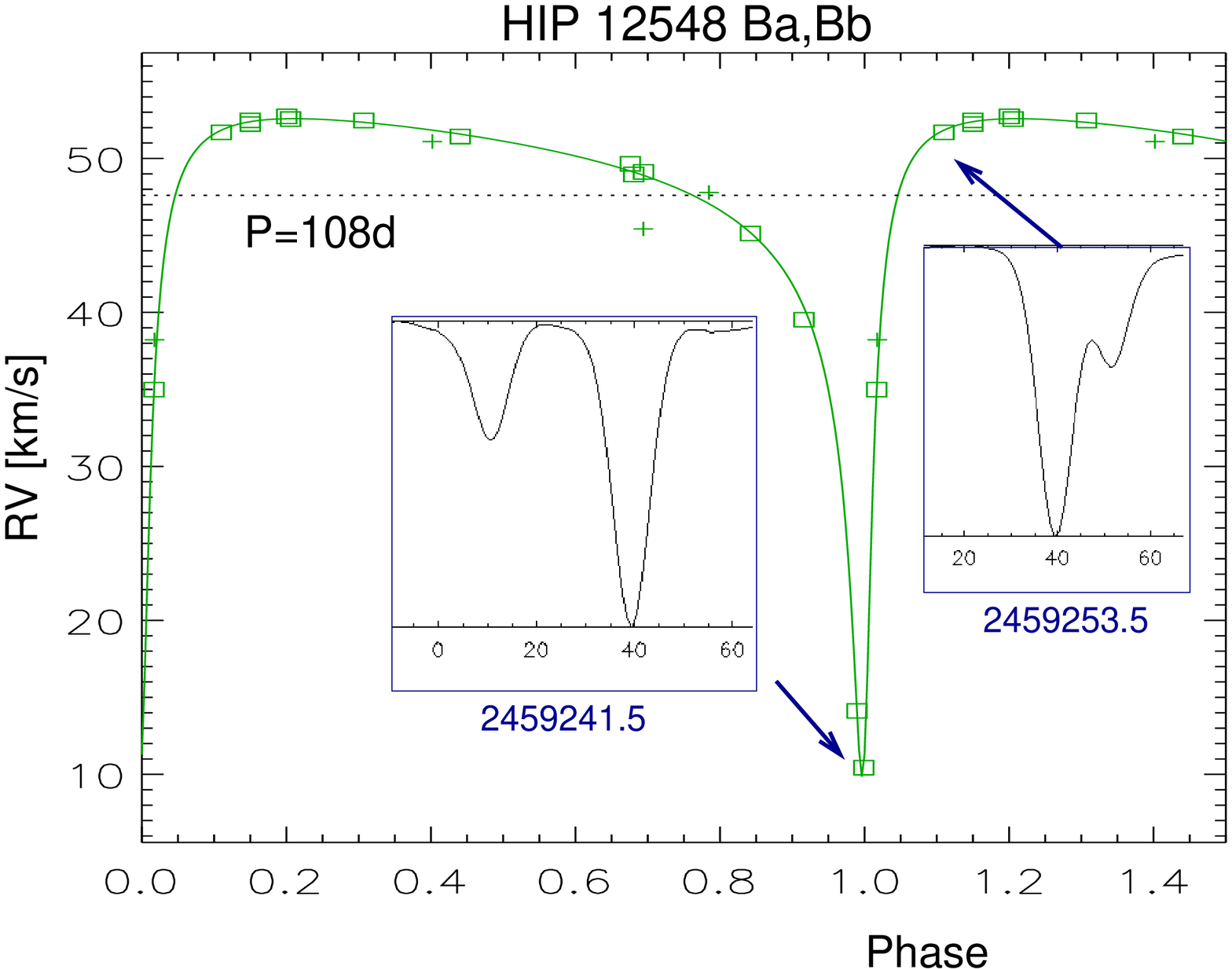}
\plotone{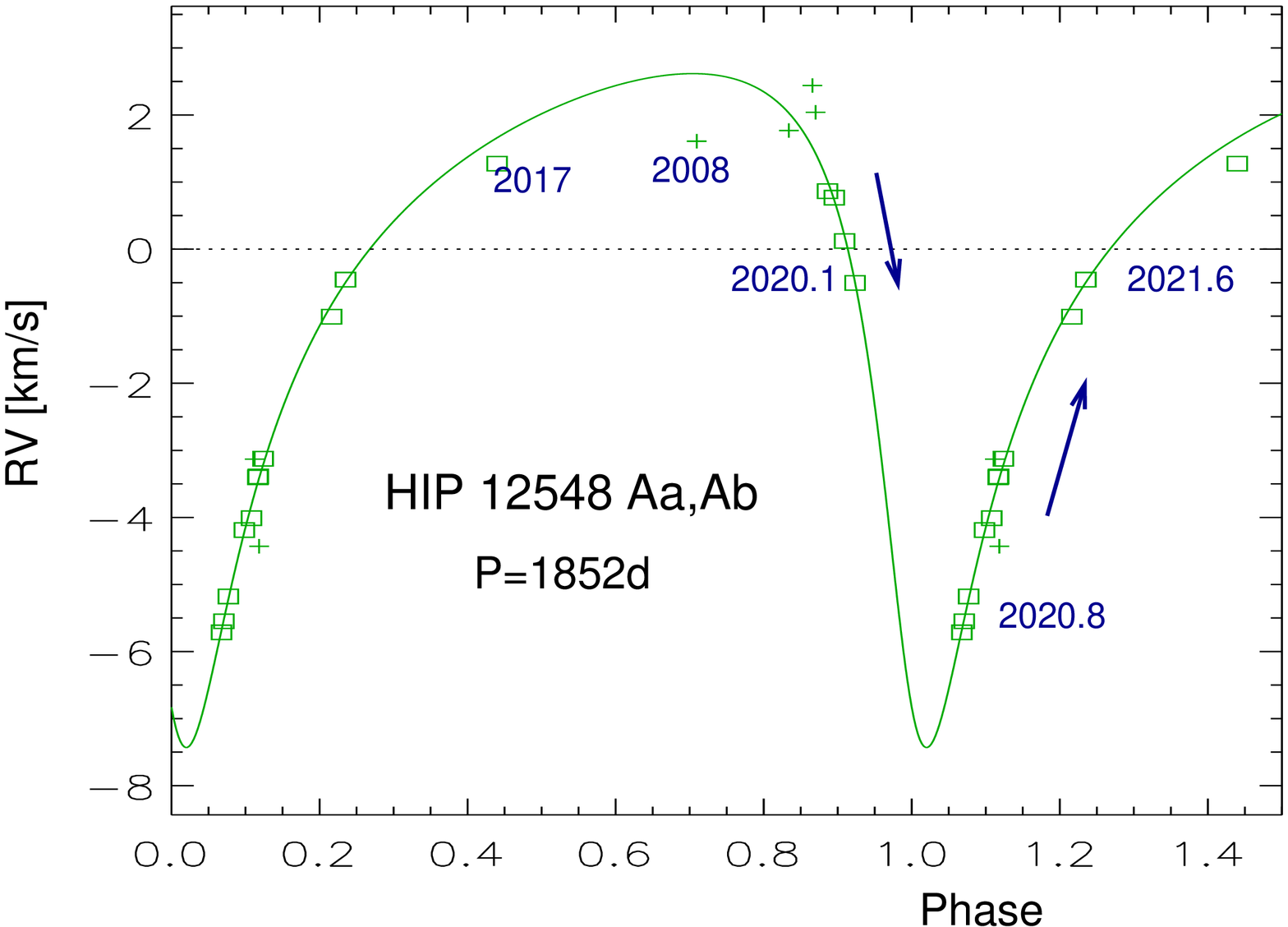}
\caption{RV curves of the components B (top) and A (bottom) of HIP
  12548. The RV measurements derived from strongly blended dips (crosses) are
  assigned large errors.
\label{fig:inner} 
}
\end{figure}

Figure~\ref{fig:inner}  illustrates  the  single-lined orbits  of  the
subsystems  Ba,Bb  and  Aa,Ab.   The   orbit  of  Ba,Bb  has  a  large
eccentricity of  0.82.  The  minimum of  its RV  curve was  covered by
regular observations  in 2021  January-February. The inserts  show the
CCFs at the RV minimum on JD 2459241.5, where the weaker dip of Ba was
located to the left of the stronger  dip of Aa.  Twelve days later, on
JD 2459253.5,  the lines of  Ba were already  seen on the  right side,
partially blending with  the lines of Aa.  The RV of Aa  on both dates
was around 40 \kms.

The   RV  curve   of   Aa,Ab  is   shown   in  the   lower  panel   of
Figure~\ref{fig:inner}. Its descending branch was covered in the fall of
2019.  CHIRON  was closed for COVID-19  pandemic in the  first half of
2020.  After  its re-opening  in 2020 October,  the increasing  RV was
measured.  The RVs  alone do  not  fully constrain  the orbit,  partly
because blended dips  of Aa and Ba result  in poor accuracy.  Reliable
dip splitting is achieved only near the periastron of the Ba,Bb orbit.

\begin{figure}[ht]
\plotone{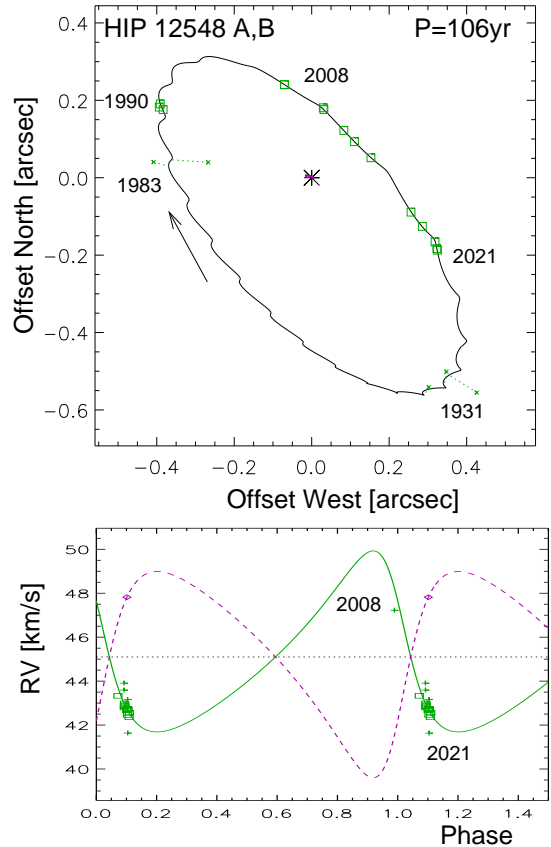}
\caption{The  outer   orbit  of  HIP  12548   A,B.   Top:  positional
  measurements (crosses denote visual micrometer measurements, squares
  are  accurate speckle  measurements,  the wavy  line  is the orbit  with
  wobble), bottom: the RV curve.
\label{fig:outer} 
}
\end{figure}

\begin{deluxetable*}{ll cccc ccc ccc}    
\tabletypesize{\scriptsize}     
\tablecaption{Visual and astrometric orbits
\label{tab:vborb}          }
\tablewidth{0pt}                                   
\tablehead{                                                                     
\colhead{HIP} & 
\colhead{System} & 
\colhead{$P$} & 
\colhead{$T$} & 
\colhead{$e$} & 
\colhead{$a$} & 
\colhead{$\Omega_{\rm A}$ } & 
\colhead{$\omega_{\rm A}$ } & 
\colhead{$i$ }  &
\colhead{$K_1$ } & 
\colhead{$K_2$ } & 
\colhead{$\gamma$ }  \\
 & & \colhead{(yr)} &
\colhead{(yr)} & &
\colhead{(arcsec)} & 
\colhead{(deg)} & 
\colhead{(deg)} & 
\colhead{(deg)} &
\colhead{(km~s$^{-1}$)} &
\colhead{(km~s$^{-1}$)} &
\colhead{(km~s$^{-1}$)} 
}
\startdata
12548 & A,B   & 106.3   & 2010.02  &0.384      &0.568       &42.4      & 63.4    & 114.7 & 4.12  & 4.69   & 45.10  \\
      &       &$\pm$4.6 &$\pm$0.19 &$\pm$0.020 &$\pm$0.012  &$\pm$0.4  &$\pm$1.6 &$\pm$0.4&$\pm$0.44  &$\pm$0.88  & fixed  \\
12548 & Aa,Ab & 5.07    &2020.477  &   0.544   &0.0117       & 70.5    &151.7    & 79.7 & 5.02 & \ldots   & \ldots \\
      &       &$\pm$0.05&$\pm$0.080&$\pm$0.040 & $\pm$0.0009 &$\pm$3.2 &$\pm$5.4  & $\pm$3.2 &$\pm$0.66    &\ldots  & \ldots  \\
76400 & A,B   & 70.0    & 2016.26  & 0.420     & 0.374       & 320.3   & 275.5    & 144.7   & 1.41         & \ldots & $-$7.28 \\ 
      &       & fixed & $\pm$0.84  &$\pm$0.010 &$\pm$0.008   &$\pm$11.5 &$\pm$21.1 &$\pm$2.9& $\pm$0.67 & \ldots &$\pm$0.71  
\enddata 
\end{deluxetable*}

The semimajor  axis of Aa,Ab is  61 mas, and the  expected photocenter
amplitude is 14\,mas.  Wobble in the  motion of A,B caused by the 5-yr
subsystem   is   detectable    in   accurate   speckle-interferometric
measurements. The  orbits of A,B  and Aa,Ab were fitted  jointly using
both the  positions and the  RVs.  Star B  was treated as  single (its
photocenter amplitude  is $\sim$2\,mas), and its  center-of-mass RV is
included as a single measurement.  The spectroscopic elements of Aa,Ab
are listed  in Table~\ref{tab:sborb}, and the  updated visual elements
of A,B and their errors are provided in Table~\ref{tab:vborb} together
with the astrometric  orbit of Aa,Ab.  The outer orbit  with wobble is
illustrated in Figure~\ref{fig:outer}.

The  first  three micrometric  measurements  of  A,B  made by  Bos  in
1931--1939 are modified here by increasing the separations by a factor
of  1.2  to  match  the  orbit (it  is  well  known  that  micrometric
measurements of separations are often  burdened by large errors, while
the position angles  are more accurate).  Unfortunately,  the pair has
not been observed  for half a century, until 1983.   The Hipparcos and
speckle  observations  in the  early  1990s  and  the SOAR  data  from
2008--2021,  covering  the  periastron,   fully  constrain  the  outer
orbit. The outer RV curve plots the systemic RV of B, a single blended
measurement in  2008 attributed to Aa,  and the CHIRON RVs  of Aa with
subtracted inner  orbit. The  weighted rms  residuals of  SOAR speckle
measurements are 0.8\,mas  in both coordinates, and  the fitted wobble
amplitude  is  11.7$\pm$0.9  mas,  in agreement  with  its  estimate.
Without the wobble, the position residuals increase to 5\,mas.  Secure
detection of  the wobble  helps to  constrain the  Aa,Ab orbit  in the
joint fit and  defines its orientation on the sky.   The RV amplitudes
of  A  and B  are  4.1  and  4.7  \kms, respectively.  However,  these
amplitudes are  poorly constrained, and  I had  to select and  fix the
systemic  velocity of  A,B  to  obtain RV  amplitudes  that match  the
estimated masses (see below).  The  angle between the angular momentum
vectors of A,B and Aa,Ab (mutual inclination) is 45\degr$\pm$4\degr.

The $V$ magnitudes of A and B are 8.15 and 9.26 mag, respectively (the
difference of  $\sim$1 mag is  confirmed at SOAR), and  their absolute
magnitudes (using the Hipparcos parallax)  correspond to the masses of
1.07  and  0.89 \msun  for  Aa  and Ba.   The  minimum  masses of  the
spectroscopic  secondaries Ab  and Bb  are 0.32  and 0.31  \msun.  The
actual  mass  of  Ab  equals  its minimum  mass  because  the  orbital
inclination  is large,  80\degr.   Contribution  of the  spectroscopic
secondaries to  the optical fluxes  is negligible and their  lines are
not seen  in the spectra.  The  total estimated mass of  the system is
thus  2.6 \msun.   The  outer orbit  and mass  sum  correspond to  the
dynamical parallax of  18.5 mas, in good agreement  with the Hipparcos
parallax of 18.6\,mas.  The estimated  masses result in the outer mass
ratio $q_{\rm AB} = 0.87$.

This quadruple system presents a tough challenge for Gaia. The satellite
did not  resolve the A,B  pair, which moves  fast on the  outer orbit.
Without  prior  knowledge  of  the  orbits,  modeling  the  non-linear
photocenter motion that includes strong wobble signals with periods of
5\,yr and 108 days will hardly be possible even in the final Gaia data
release. Blended spectra preclude the Gaia RV measurements.

\subsection{HIP 21079 (Triple)}

\begin{figure}
\epsscale{1.0}
\plotone{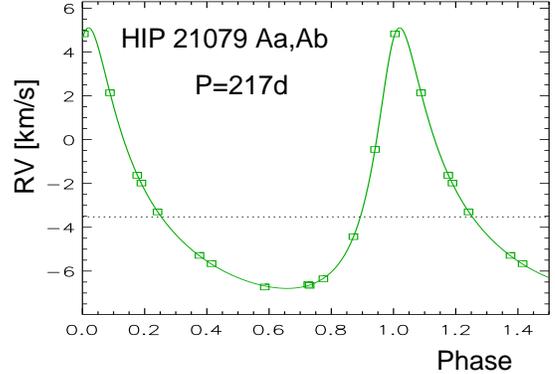}
\caption{The RV curve of HIP 21079 Aa,Ab.
\label{fig:21079}
}
\end{figure}

This object  is a  nearby (45\,pc)  solar-type star.   Its astrometric
acceleration  in   Hipparcos  \citep{MK05}   prompted  high-resolution
observations at  Gemini, detecting  a faint  companion B  at 1\farcs62
separation  with  $\Delta  K  =  1.91$  and  $\Delta  H  =  2.14$  mag
\citep{Tok2012}. This pair,  TOK~208, was observed at  SOAR at similar
position and with $\Delta I =  3.52$ mag. Its components have separate
entries  in  Gaia, adding  $\Delta  G  =  3.63$  mag to  the  relative
photometry. 

The absolute magnitude of the primary star A matches its spectral type
G3V  and  corresponds  to  the  mass  of  0.98  \msun.   According  to
\citet{Bensby2018}, the  effective temperature  of A  is 5792$\pm$48K,
while [Fe/H]=$-$0.11$\pm$0.05.  They detected  the lithium line in the
spectrum and  determined the  Li abundance  of 2.50.   The multi-color
differential  photometry of  A,B is  consistent with  B being  an 0.55
\msun star.  The  projected separation of A,B implies a  period on the
order  of  500  yr,  in   agreement  with  the  slow  observed  motion
(increasing  separation  at  constant angle).  The  small  astrometric
acceleration revealed  by the $\Delta  \mu$ could be produced  by this
pair.

\citet{N04} found that the RV of HIP 21079 varies with an amplitude of
3.8 \kms.  The star was placed on  the CHIRON program in 2015 with the
aim of  determining its spectroscopic  orbit. The spectra show  a weak
lithium line  and indicate a  moderate rotation of $  V \sin i  = 5.4$
\kms.  A  single-lined orbit of  Aa,Ab with  $P=217$ days is  shown in
Figure~\ref{fig:21079}.  The  residuals are  0.024 \kms ~owing  to the
narrow and high-contrast  CCF dip.  The minimum mass of Ab  is 0.16 \msun,
and no  traces of its  lines are seen  in the spectra.   The semimajor
axis of  Aa,Ab is 16.2\,mas,  and the photocenter amplitude  should be
2.3\,mas. This system is a  planetary-type hierarchy with two low-mass
companions orbiting the central solar-type star.

\subsection{HIP 24320 (Quadruple)}

\begin{figure}
\plotone{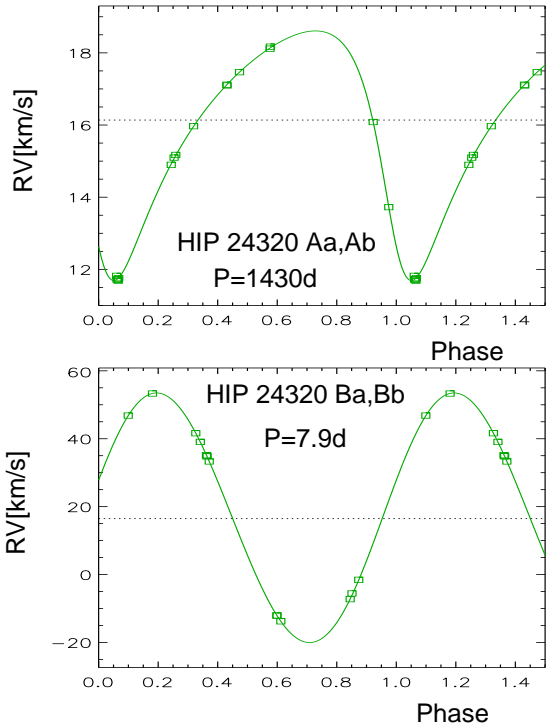}
\caption{The RV curves of single-lined spectroscopic subsystems Aa,Ab
  (top) and Ba,Bb (bottom) of the HIP 24320 quadruple.
\label{fig:24320}
}
\end{figure}

This is a quadruple system of 2+2 hierarchy. The outer visual pair A,B
has a separation  of 4\farcs63 and a magnitude  difference $\Delta V =
1.26$,  $\Delta G  = 1.15$  mag. It  is cataloged  under the  name of
HJ~3743, having been  discovered by J.~Herschel in 1835.  Stars A and B
have   individual   astrometry  in   Gaia,   proving  their   physical
relation. The estimated period of A,B is 3 kyr.

\citet{N04} indicate that the RV is variable, but do not specify which
component  of the  visual  pair was  observed.   Both components  were
observed with CHIRON,  and both have variable  RVs. Their single-lined
orbits  with  periods  of  1430   and  7.9  days  are  illustrated  in
Figure~\ref{fig:24320}.  Despite the short  period, the Ba,Bb pair has
a measurable  eccentricity of  0.024. Adopting the  masses of  1.0 and
0.82 \msun  for Aa and  Ba, appropriate for their  absolute magnitudes
and spectral types G3V and K2V,  the minimum masses of the secondaries
are 0.19 and 0.39 \msun, respectively.

The astrometry  of A in  Gaia EDR3 is likely  distorted by the  3.9 yr
subsystem: the  estimated astrometric wobble  amplitude is at  least 7
mas.    The  acceleration   of  A   is  evidenced   by  $\Delta   \mu$
\citep{Brandt2018}  and by  RUWE=15.0. Future  Gaia data  releases are
expected to contain the astrometric orbit of Aa,Ab.

\subsection{HIP 27970 (Triple)}

\begin{figure}
\plotone{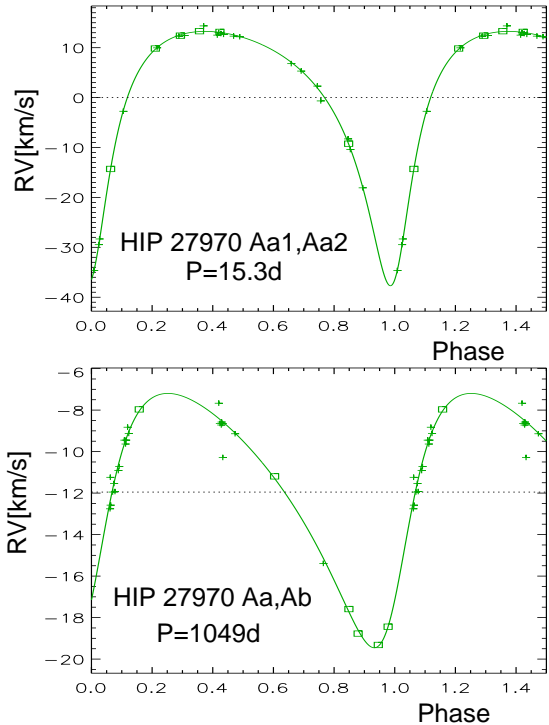}
\caption{The RV curves of the inner (top) and outer (bottom) orbits in
  HIP 27970. Crosses correspond to the RVs from \citet{Gorynya2018},
  squares --- to the CHIRON RVs.
\label{fig:27970}
}
\end{figure}

HIP 27970 (HD  39899, F7V, $V=7.73$ mag) is a  spectroscopic triple in
the solar  neighborhood (distance  60\,pc).  No visual  companions are
listed in  the WDS and  none are found  in Gaia within  2\arcmin. The
Gaia  EDR3  parallax  of  15.78\,mas  is  biased  by  the  astrometric
acceleration,  but  nevertheless  agrees,   within  errors,  with  the
Hipparcos parallax  of 16.0$\pm$1.3 mas. The  astrometric acceleration
was detected by Hipparcos and confirmed by \citet{Brandt2018}.

The RV variability was reported  by \citet{N04}. The star was observed
in   2012--2016  with   the  correlation   radial-velocity   meter  by
\citet{Gorynya2018}, who determined a single-lined spectroscopic orbit
with $P=15.3$  days. Residuals from  this orbit suggested  presence of
another companion  with a  longer period that  would also  explain the
acceleration.  Additional  observations with  CHIRON made in  2020 and
2021 are used  here to determine the orbit  of this tertiary companion
with $P=1049$  days (2.9 yr).  Both orbits  were fitted simultaneously
using {\tt  ORBIT3} \citep{TL2017}.  The  errors of the  published RVs
were  adjusted by adding  quadratically the  instrumental error  of 0.3
\kms;  the two  most deviant  RVs were  given larger  errors.  Another
strongly deviant RV  from that paper corresponds to  the periastron of
the outer orbit, explaining the apparent disagreement with the initial
orbit.  The  errors of the CHIRON  RVs of 0.2 \kms  were adopted.  The
weighted rms residuals are 0.22 \kms.
Figure~\ref{fig:27970}  shows  both RV  curves.  The  inner and  outer
eccentricities are  0.504 and 0.359,  respectively, and the  center of
mass velocity is $-11.95$ \kms.

I adopt the mass of 1.25 \msun for the primary component Aa1, estimated
from its  absolute magnitude and  matching  the spectral type F7V  and the
$V-K$ color.   The orbits give minimum  masses of 0.35  and 0.44 \msun
for Aa2 and Ab, respectively. The semimajor axis of Aa,Ab is 40 mas.

Both  orbits have  substantial inclinations,  otherwise the  secondary
stars should be  more massive and then would become  detectable in the
spectra and add the flux, moving the object above the main sequence in
the  color-magnitude  diagram  (in  fact  it  is  right  on  the  main
sequence). If the actual masses are close to their minimum values, the
astrometric semimajor  axis of the outer  orbit is a 0.22  fraction of
the full axis, i.e.  8.6 mas.  The PM anomaly $\Delta \mu$ measured by
\citet{Brandt2018} is $(-10.7, -6.0)$ mas~yr$^{-1}$, its modulus is 12.3
mas~yr$^{-1}$.  The outer orbit  with an inclination of 90\degr ~predicts
a photo-center  motion of  18.4 mas~yr$^{-1}$ over  the Gaia  DR2 time
span, in qualitative agreement with  this crude estimate. The position
angle  of the  outer node  should be  around 60$^\circ$  to match  the
$\Delta \mu$ direction.

The  RV of  the center  of  mass and  the long-term  PM determined  by
\citet{Brandt2018}  correspond to  the  Galactic  velocity $(U,V,W)  =
(17.3,  2.4,  -7.4)$  \kms.   The  object thus  belongs  to  the  disk
population, typical of  the solar neighborhood. A  very strong lithium
line is present in the CHIRON spectra, despite the slow rotation of $V
\sin i = 2.8$ \kms.

\subsection{HIP 34212 (Triple)}

\begin{figure}
\epsscale{1.0}
\plotone{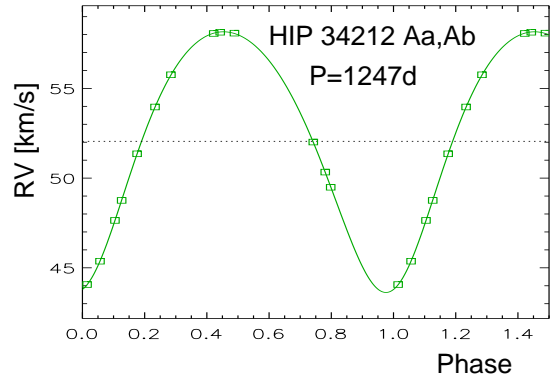}
\caption{RV curve  of HIP 34212 Aa,Ab. 
\label{fig:34212}
}
\end{figure}

This is  yet another  typical solar-type  triple. The  outer 11\arcsec
~pair A,B (TOK~114)  has been identified in  \citep{Tok2011} and
confirmed as  CPM by Gaia.  The  EDR3 parallax of B  ($V=16.9$ mag) is
16.55\,mas (distance 60.4\,pc), its estimated  mass is only 0.2 \msun.
The EDR3 parallax of A, 14.39\,mas,  is biased by the subsystem Aa,Ab,
which has  been revealed  by  variable RV  \citep{N04} and  by 
astrometric  acceleration  \citep{MK05}; the  large  RUWE  of 20.7  is
caused by the subsystem.

Star A  was observed with CHIRON  since 2015. The  spectrum has single
lines slightly  broadened by  rotation ($V \sin  i = 5.3$  \kms).  The
orbit    of    Aa,Ab   with    $P=3.4$    yr    is   illustrated    in
Figure~\ref{fig:34212}; the  residuals are  only 0.015 \kms.   Star Aa
(F8V) has an estimated mass of  1.24 \msun, and the minimum mass of Ab
is 0.53  \msun.  Lines of Ab  are not detectable in  the spectra.  The
semimajor axis of Aa,Ab is 45 mas, the photocenter axis is 13.6\,mas.

\subsection{HIP 56282 (Quadruple)}

\begin{figure}
\epsscale{1.0}
\plotone{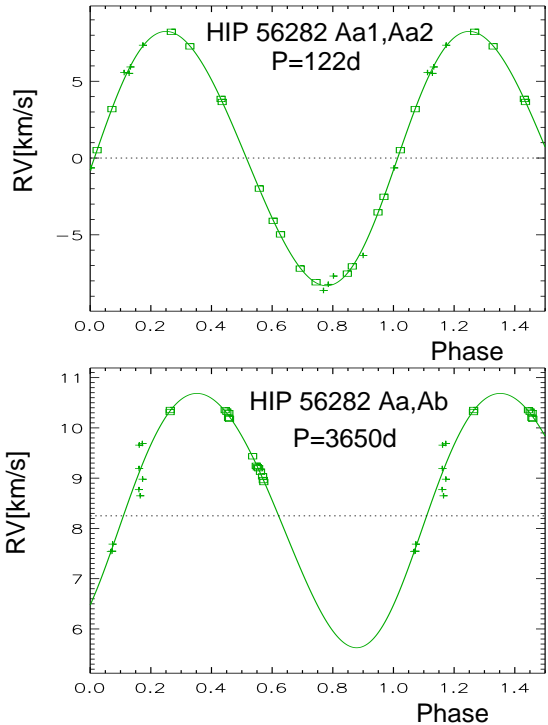}
\caption{RV curves  of HIP 56282: top -- inner orbit, bottom --
  intermediate orbit. Crosses denote the less accurate RVs from
  \citet{Gorynya2018}. The intermediate orbit is still tentative.
\label{fig:56282}
}
\end{figure}

This is  a solar-type F8V  star located at  61 pc distance.  The outer
15\farcs9 pair A,B was identified  in \citep{Tok2011} and confirmed as
physical  by the  Gaia  astrometry.   Star B  ($V=16.23$  mag) has  an
estimated  mass of  0.30 \msun,  and its  EDR3 parallax  of 16.28  mas
defines the distance.   Star A contains  a subsystem, as follows  from the
variable RV  \citep{N04} and the acceleration  \citep{MK05}. Gaia EDR3
gives a  biased parallax of 15.13\,mas  with a large RUWE  of 6.4. The
long-term PM of A is in excellent  agreement with the PM of B, but its
short-term PM  measured by  Gaia is different  owing to  the subsystem
\citep{Brandt2018}.

The RV of A has been monitored by \citet{Gorynya2018} since 2016 using
a  correlation radial-velocity meter  to determine  the period  of the
subsystem.  The  data were not  sufficient for the  orbit calculation,
and  it was  suspected  that star  A  is triple.   This hypothesis  is
confirmed  by the additional  CHIRON data  presented here.   The inner
period  of 122 days  is clearly  identified, superposed  on a  slow RV
variation.  I designate  the innermost  subsystem as  Aa1,Aa2  and the
intermediate one as Aa,Ab. The combined RV data set does not yet fully
cover the Aa,Ab orbit, and I adopt a preliminary period of 10 yr while
fitting    both    orbits    simultaneously   using    {\tt    ORBIT3}
(Figure~\ref{fig:56282}). With a  coverage of 1841 days, substantially
shorter outer  periods are excluded.  Longer periods, e.g.  15--20 yr,
are possible, but they require a larger RV amplitude and a larger mass
of Ab (no trace of Ab is seen in the CCF). So, the orbit of Aa,Ab
proposed here is  hypothetical, but plausible. 

The  minimum  masses   of  Aa2  and  Ab  are  0.24   and  0.25  \msun,
respectively,  if the  mass of  1.14 \msun  is adopted  for Aa1.   The
semimajor  axis of  Aa,Ab  is  89 mas,  and  the  photocenter axis  is
13.6\,mas.  The Aa,Ab orbit predicts a  $\Delta \mu$ of the same order
as observed, about 5~mas~yr$^{-1}$  (the orbit orientation is unknown,
precluding a more detailed comparison). 

The innermost  orbit is almost circular, and  the middle orbit  also seems to
have a small  or zero eccentricity. Thus, the  3+1 quadruple HIP~56282
resembles a planetary system by  its quasi-circular orbits and the low
masses of  all three companions relative  to the main  star, making it
similar to HD~91962 \citep{HD91962}.

Star  A has  a  strong  lithium line  in  its  spectrum, although  the
projected   rotation   is    slow,   $V   \sin   i    =   3.6$   \kms.
\citet{Armstrong2015}  found a  flux modulation  with an  amplitude of
3.17  per  cent  and  no  apparent  period  in  the  Kepler  K2  data.
\citet{Chaplin2015} also used the K2 photometry to detect astroseismic
signals  from solar-type  stars;  none was  found  in HIP~56282.  They
determined stellar parameters from  a high-resolution spectrum: $T_e =
6009$\,K, $V \sin i =4.7$ \kms, and [Fe/H] = $-$0.26.

\subsection{HIP 57860 (Triple)}

\begin{figure}
\epsscale{1.0}
\plotone{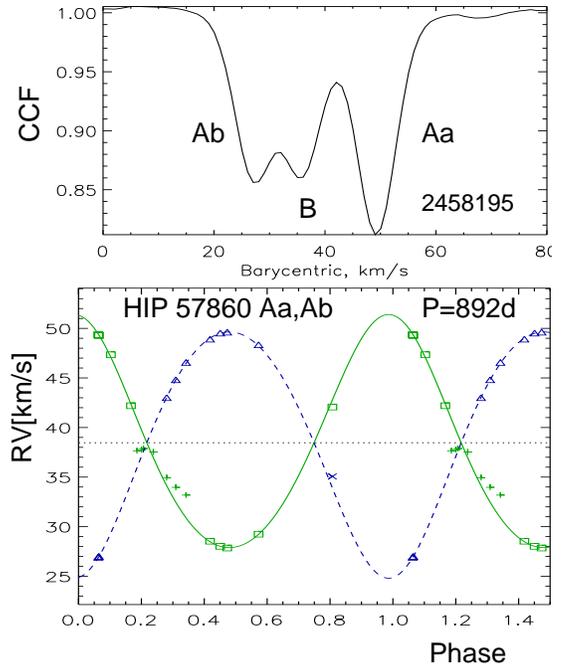}
\caption{CCF (top)  and RV curve (bottom) of HIP 57860 Aa,Ab. 
\label{fig:57860}
}
\end{figure}

The spectra of this G5V star  are triple-lined. The outer pair A,B has
been known since 1897 and  is designated as SEE~137. The separation is
1\farcs63.  Differential  speckle photometry  at  SOAR yields  $\Delta
I_{\rm AB} = 0.40$, $\Delta y_{\rm  AB} = 0.25$ mag, Gaia gives $\Delta
G_{\rm AB}  = 0.25$ mag.  The period of  A,B is estimated  as $\sim$900
yr. Gaia measured matching parallaxes and PMs of both components.

Observations with  CHIRON lead to  a double-lined orbit of  Aa,Ab with
$P=892$ days  and a  small eccentricity  (Figure~\ref{fig:57860}). The
spectroscopic masses  $M \sin^3 i$  of Aa and  Ab are 0.66  ans 0.62
\msun, the masses  estimated from the absolute magnitudes  are 0.97 and
0.92 \msun, suggesting an inclination of 62\degr. The inner mass
ratio  is 0.95,  the ratio  of the  dip areas  implies $\Delta  V_{\rm
  Aa,Ab} = 0.30$ mag.

Similar luminosities of Aa and Ab  and the estimated semimajor axis of
the inner orbit,  28.7\,mas, make this pair  potentially resolvable by
speckle. However, seven observations at SOAR  made so far only hint at
marginal  resolution of  the inner  subsystem, while  the A,B  pair is
measured.

The RV of the visual secondary B measured from the triple-lined
spectra is 36.01 \kms with the rms scatter of 0.07 \kms. The estimated
mass of B is 1.04 \msun, so all three stars in this system are
similar. Their lines are narrow, indicating slow axial rotation.

\subsection{HD 108938 (Triple)}

\begin{figure}
\epsscale{1.0}
\plotone{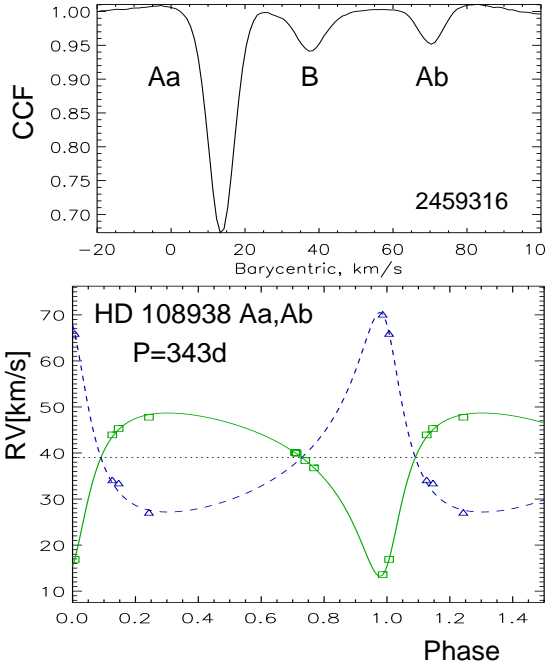}
\caption{CCF (top)  and RV curve (bottom) of HD 108938.
\label{fig:108938}
}
\end{figure}

This  system is  an analogue  of the  previous one,  i.e.  a classical
visual binary  with a triple-lined spectrum. The  outer 1\farcs23 pair
A,B  has been  known since  1934 and  is designated  as  RST~2802. The
magnitude differences  are $\Delta I_{\rm  AB} = 1.92$ mag  (SOAR) and
$\Delta G_{\rm AB} = 1.58$ mag (Gaia), and the estimated period of A,B
is 1.4 kyr.  Gaia EDR3 gives a biased parallax  of 5.82\,mas for A
(RUWE=7.1), so the parallax of B, 6.35\,mas, is adopted.

The orbit of  the subsystem Aa,Ab with $P=343$ days  is illustrated in
Figure~\ref{fig:108938}.   The mass  ratio is  0.82 and  the magnitude
difference estimated from the ratio of the dip areas is $\Delta V_{\rm
  Aa,Ab} =  2.2$ mag.  The estimated  semimajor axis of Aa,Ab  is only
8.4\,mas, making  resolution of the  subsystem by speckle  an unlikely
prospect. However,  its astrometric  orbit is  expected in  the future
Gaia releases. Comparison of $M \sin^3 i$ with the estimated masses of
Aa and  Ab (1.43 and 1.16  \msun) leads to an  inclination of 53\degr.
Stars Aa, Ab, and B are  less similar between themselves in comparison
to  HIP~57860, but  also  rotate  slowly and  do  not have  detectable
lithium lines in  their spectra. The estimated masses  are larger than
expected for  the spectral type, so  the primary might be  leaving the
main sequence.

\subsection{HIP 76400 (Quadruple)}

\begin{figure}
\epsscale{1.0}
\plotone{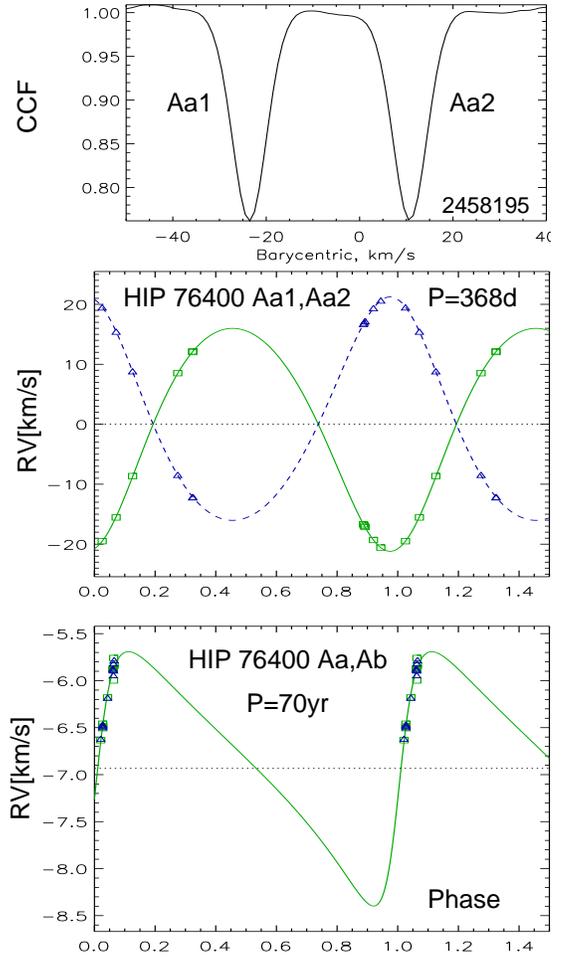}
\caption{CCF (top) and RV curve (middle) of HIP 76400 Aa1,Aa2 with the
  trend subtracted.   The lower plot shows  the RV curve  in the 70-yr
  visual orbit of Aa,Ab.
\label{fig:76400}
}
\end{figure}

This is a quadruple system of  3+1 hierarchy.  A faint ($V=17.82$ mag)
CPM  companion  B at  80\farcs4  separation  was discovered  from  the
ground-based data and confirmed by Gaia.  The estimated period of this
pair  is $\sim$230  kyr.  Astrometric  acceleration of  the main  star
\citep{MK05}  prompted its  high-resolution  observations  at SOAR  in
2014,  revealing another  companion  Ab at  0\farcs19 separation  with
$\Delta  I_{\rm  Aa,Ab}   =  3.84$  mag.   This   companion  has  been
independently found in 2011 by \citet{Horch2017}. A preliminary visual
orbit of this  pair with $P=70$ yr  and a 0\farcs34 axis  based on the
140\degr ~arc  observed in  2011--2021 was  determined by  the author.
Motion  of the  photocenter predicted  by  this orbit  in 1991.25  and
2015.5 matches  the actual $\Delta  \mu$, strengthening the  orbit and
suggesting a mass ratio of 0.25. The masses of B and Ab estimated from
their absolute magnitudes are 0.17 and 0.61 \msun respectively.

\citet{N04} noted  that this star  had double lines.   Monitoring with
CHIRON leads to the  orbit of the inner pair Aa1,Aa2  with a period of
one year shown  in Figure~\ref{fig:76400}.  The mass  ratio of Aa1,Aa2
equals  one  within error,  the  CCF  dips  have  the same  width  and
contrast,  so  this is  a  perfect  twin.   No astrometric  wobble  is
expected, therefore the Gaia EDR3  parallax can be trusted (indeed, it
equals  the parallax  of  B within  errors).  Masses  of  Aa1 and  Aa2
estimated   from  the   absolute  magnitudes   are  1.03   \msun,  the
spectroscopic  masses   are  0.96  \msun,  hence   the  inner  orbital
inclination is  $\sim$77\degr ~or  $\sim$103\degr.  To account  for the
small  RV trend  caused by  motion in  the visual  orbit, I  fitted it
jointly with the  inner orbit using {\tt ORBIT3}. As  the observed arc
does not yet constrain  the period, I fixed it to  70\,yr to match the
EDR3 parallax  and the estimated  mass sum,  2.67 \msun. The  outer RV
amplitude of  1.4 \kms reproduces  the observed trend of  the systemic
velocity   of   Aa1,Aa2,   as   shown    in   the   lower   panel   of
Figure~\ref{fig:76400}.  The  elements of  this tentative  Aa,Ab orbit
are  given  in  table~\ref{tab:vborb};  the  inclination  of  144\degr
~indicates that the two orbits are not coplanar.

\subsection{HIP 76816 (Quadruple)}

\begin{figure}
\epsscale{1.0}
\plotone{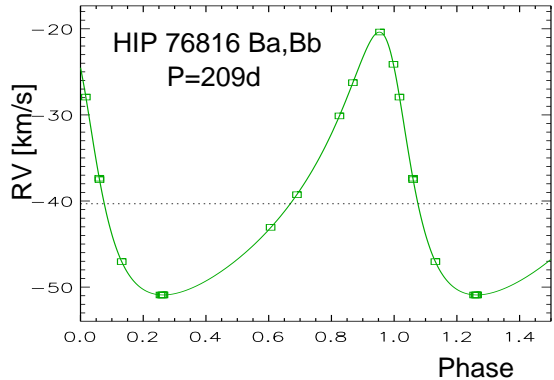}
\caption{RV curve of HIP 76816 Ba,Bb.
\label{fig:76816}
}
\end{figure}

This quadruple  system has  been presented in  Paper 4 of  this series
\citep{chiron4}, where  the double-lined orbit of  Aa,Ab with $P=6.95$
days  has  been determined.   The  outer  pair  A,B (ADS~9743)  has  a
separation of 5\farcs5. Gaia  EDR3 gives concordant parallaxes and PMs
of A and B, placing the system  at a distance of 308\,pc; the estimated
period of A,B  is 32 kyr. Both  A and B are evolved,   well above
the main sequence.

Early CHIRON observations gave a strong evidence that the RV of star B
is variable. Continued monitoring of this component leads to its orbit
with $P=208.9$ days presented  in Figure~\ref{fig:76816}.  One RV from
\citet{Desidera2006}  is used  to increase  the period  accuracy.  The
residuals are very small, 0.016 \kms. The estimated mass of Ba is 1.67
\msun, and the  minimum mass of Bb is then  0.68 \msun. For reference,
the masses  of Aa and  Ab are 1.72  and 1.27 \msun,  respectively. The
semimajor axes of the inner subsystems  are 0.3 and 2.6\,mas, none has
been resolved  by speckle interferometry at  SOAR, and only B  gives a
marginal evidence  of increased astrometric  noise in Gaia (RUWE  of A
and B are 1.1 and 1.6, respectively).

\subsection{HIP 81394+81395 (Triple)}

\begin{figure}
\epsscale{1.0}
\plotone{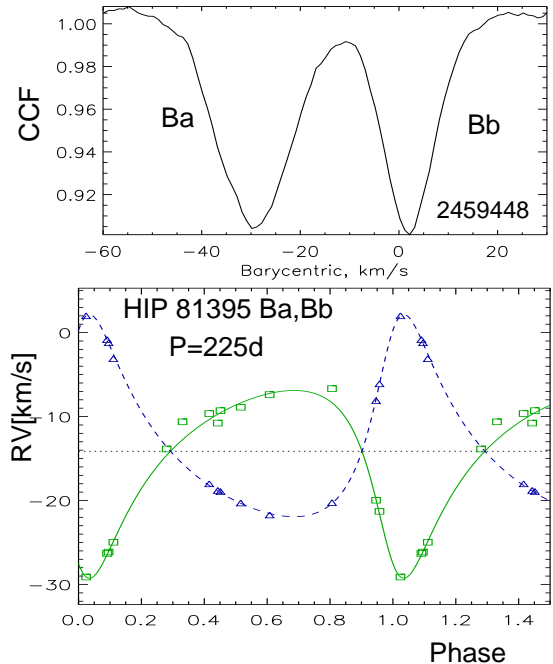}
\caption{CCF (top)  and RV curve (bottom) of HIP 81395 Ba,Bb, the visual
  companion of HIP~81394.
\label{fig:81394}
}
\end{figure}

The  outer 10\farcs9  pair of  this triple  system, HJ~4862,  has been
first measured  by J.~Herschel in  1835.  Its components  have matching
parallaxes and PMs  in Gaia EDR3, without  astrometric acceleration or
increased astrometric noise (RUWE of  1.3 and 1.6).  The small $\Delta
\mu$  of these  stars  computed by  \citet{Brandt2018} are  marginally
significant, but one has to bear in mind that the Hipparcos astrometry
of  such  visual   binaries  has  systematic  errors   caused  by  the
satellite's measuring system design.

Double lines  in star B were noted  by \citet{Desidera2006}, prompting
its  monitoring with  CHIRON since  2017.  The orbit  of Ba,Bb  with
$P=225$ days is presented in Figure~\ref{fig:81394}.  The CCF dips are
broad  and  of  low  contrast,  while the  RV  amplitudes  are  small.
Reliable  fitting  of  the  double  dips is  possible  only  near  the
periastron; in other orbital phases,  the RVs derived from the blended
dips are uncertain, and they were  given low weights in the orbit fit,
which also uses the RVs from Desidera et al.

The  small  RV amplitudes  of  Ba  and Bb  are  explained  by the  low
inclination of $\sim$25\degr, estimated by comparing the expected mass
of Ba,  1.35 \msun, with $M_1  \sin^3 i = 0.11$  \msun.  The semimajor
axis  of  Ba,Bb   is  5\,mas.  This  star  was   observed  by  speckle
interferometry  at SOAR  and  found  unresolved.  Stars  A  and B  are
evolved (above the main sequence), and A is slightly redder than B.

\subsection{HIP 98294 + 98278 (Triple)}

\begin{figure}
\epsscale{1.0}
\plotone{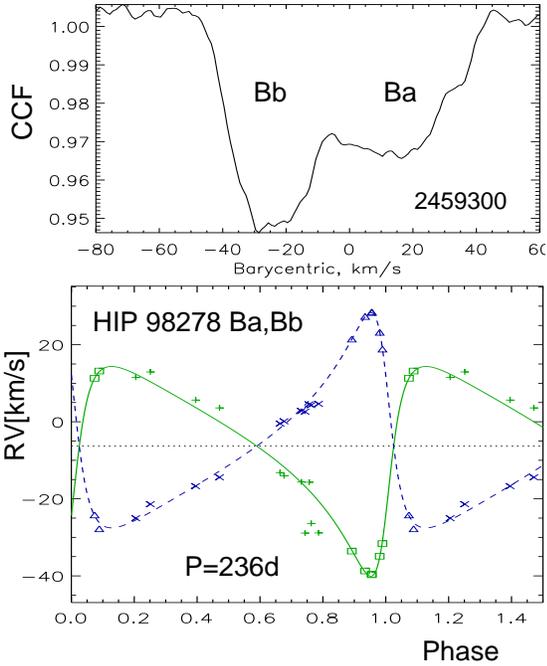}
\caption{CCF (top)  and RV curve (bottom) of HIP 98278 Ba,Bb, the visual
  companion of HIP~98294. Crosses denote inaccurate RVs deduced from
  blended dips. 
\label{fig:98294}
}
\end{figure}

The A9IV star HIP 98294 and the F6V star HIP 98278, at 80\farcs6 from
each other, have common PMs and parallaxes and form a physical pair
DUN~229 with an estimated period of $\sim$400 kyr. 

\citet{N04} noted  that star  B has double  lines. The CCFs  of CHIRON
spectra  (Figure~\ref{fig:98294}) have  shallow, broad,  and partially
blended   dips  that   are  more   or  less   separated   only  around
periastron.  The RVs  of blended  dips were  determined by  fixing the
amplitude of the wide dip belonging  to Ba and are given a lower weight in
the  orbit  fit.  The  residuals   are  unusually  large,  as  can  be
appreciated in  the RV curve.  

The ratio  of the dip  areas gives a  crude estimate of  the magnitude
difference  $\Delta   V_{\rm  Ba,Bb}  \approx  0.8$   mag,  hence  the
individual magnitudes  are 8.6 and  9.4, and  the masses are  1.25 and
1.08 \msun  (ratio 0.86).  The spectroscopic masses  $M \sin^3  i$ are
only slightly less, so the inclination is high, $i_{\rm Ba,Bb} \approx
76^\circ$;  the spectroscopically  measured  mass ratio  is 0.97.  The
semimajor axis of Ba,Bb is 10.5\,mas.  A small $\Delta \mu$ of star B,
0.67  mas~yr$^{-1}$ \citep{Brandt2018},  is presumably  caused by  the
wobble of the photocenter.

The  double-lined  pair B  is  located  above  the main  sequence,  as
expected, while  A is not  elevated (unevolved), despite  its spectral
type  A9IV.   Lithium lines  are  detectable  in  the spectrum  of  B,
confirming  relative youth of this system.  Assuming that A and B are
single stars,  \citet{Bochanski2018} estimated  an age of  $\sim$1 Gyr
and  masses of  1.49 and  1.27 \msun  for both  components  by fitting
models   to  the   multi-color   photometry.  Typically,   researchers
interested in wide binaries tend to  ignore the fact that many of them
contain inner subsystems.

\subsection{HIP 100420 (Triple)}

\begin{figure}
\epsscale{1.0}
\plotone{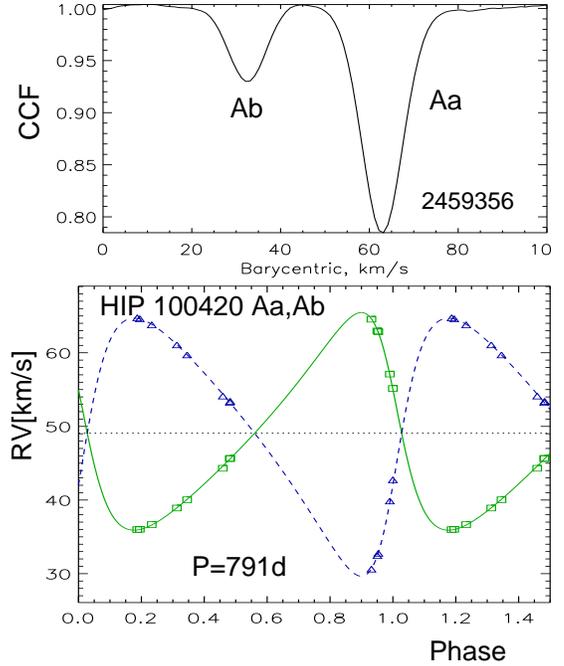}
\caption{CCF (top)  and RV curve (bottom) of HIP 100420.
\label{fig:100420}
}
\end{figure}

This  is a typical  triple system,  where double  lines were  found by
\citet{N04} in  a visual  binary HJ~5189. Its  components A and  B are
separated  by 7\farcs518 and  have a  magnitude difference  of $\Delta
G_{\rm  A,B} =  0.93$ mag.  The period  of A,B  is $\sim$25  kyr. The Gaia
astrometry yields common PMs and parallaxes of the components.

Observations of component A with  CHIRON confirm is double-lined nature
and   lead   to  the   orbit   with   $P=790.6$  days   presented   in
Figure~\ref{fig:100420}.  The  mass ratio of  the inner pair  Aa,Ab is
0.84, and the magnitude difference between  Aa and Ab deduced from the
dip areas  is 1.3  mag.  The  mass of Aa  estimated from  the absolute
magnitude is  1.53 \msun, larger  than expected  for an F6V  star, and
this component  is slightly  evolved (above  the main  sequence).  The
spectroscopic masses of Aa and Ab, $M \sin^3 i$, are 1.20 and 1.00
\msun, and the estimated inclination is $i_{\rm Aa,Ab} = 67\degr$. 

The astrometric acceleration of A detected by \citet{Brandt2018} is
likely caused by the 2-yr spectroscopic subsystem. Its semimajor axis
is about 13\,mas. The center of mass RV of A, 49.1 \kms, matches the
RV of B, measured  from three CHIRON spectra and also by Gaia. It is
thus unlikely that star B hosts a subsystem.

\section{Discussion}
\label{sec:disc}

\begin{figure}
\epsscale{1.0}
\plotone{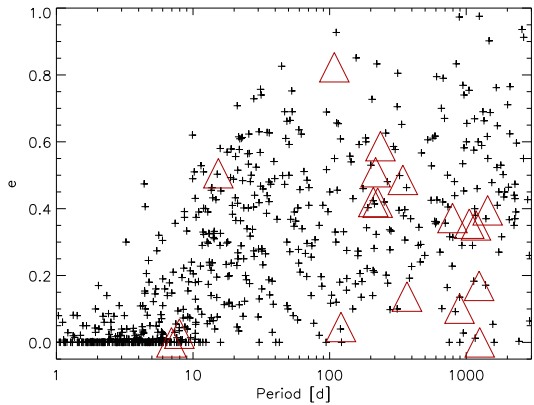}
\caption{Periods and eccentricities of inner subsystems with
  solar-type components  in the MSC (crosses). The orbits
  determined here are plotted by large red triangles. 
\label{fig:pe}
}
\end{figure}

Figure~\ref{fig:pe}  contains  a  period-eccentricity plot  for  inner
subsystems in hierarchies,  updating a similar plot from  paper 6. One
can  appreciate how  the accumulation  of CHIRON  data during  several
years results  in orbits with periods  longer than 100 days.  The time
span  needed  to  compute  such  orbits  must  exceed  their  periods,
sometimes  substantially.   Annual   visibility  cycles  can  restrict
coverage of  certain orbital  phases.  This is  particularly important
for the eccentric  orbits: a missed periastron must  be re-observed in
the next cycle(s).   The closure of CHIRON in 2020  due to the COVD-19
pandemic also had a negative impact  on this program (note the missing
coverage  of the  periastron of  a 5-yr  orbit in  the lower  panel of
Figure~\ref{fig:inner}).  Several preliminary orbits with long periods
determined with CHIRON still await additional coverage before becoming
publishable (e.g.   HIP 41171B,  967 days; HIP  75663A, 623  days; HIP
78163B, 2082 days). The next (and last) paper of this series will
present the long-period orbits and the RVs of stars where orbit
determination was not possible.  

The goal  of this program is  to obtain a  complete orbital statistics
for  subsystems with  periods  below 1000  days  in nearby  solar-type
hierarchies. The  difficulty of computing such orbits  means that many
still remain unknown, especially  those with large eccentricities. Two
orbits presented here occupy the  upper envelope of the $P-e$ relation
in  Figure~\ref{fig:pe},  namely   HIP  27970  Aa1,Aa2  ($P=15$  days,
$e=0.51$)  and   HIP  12548  Ba,Bb  ($P=108$   days,  $e=0.82$).  Both
subsystems have  small mass ratios and outer  companions on relatively
compact  orbits. The origin of binary eccentricities is still debated,
there are no models of their distribution. Hopefully, the unbiased
statistics  delivered by this program will help to develop such
models. The diverse architecture of stellar hierarchies and its
relation to their formation are discussed further in
\citet{Mult2021}. 

The Gaia mission provides a wealth of information on stars, including
multiples. This work highlights the caveats of the Gaia astrometry related
to multiplicity, namely the missing or biased data. In many cases the
future Gaia data releases will rectify this  by including orbital
motion in the astrometric model. However, the presence of additional
nearby companions and motion on several orbits will still present
problems. Meanwhile, an elevated RUWE remains a useful diagnostic of
potentially problematic Gaia astrometry, as demonstrated here.

\begin{acknowledgments} 

I thank operators of the 1.5-m telescope for executing observations of
this  program  and  the   SMARTS  team  for  scheduling  and  pipeline
processing  and the referee, K.~Fuhrmann, for  careful check of the manuscript.

The research was funded by the NSF's NOIRLab.
This work  used the  SIMBAD service operated  by Centre  des Donn\'ees
Stellaires  (Strasbourg, France),  bibliographic  references from  the
Astrophysics Data  System maintained  by SAO/NASA, and  the Washington
Double  Star Catalog  maintained  at USNO.  This  paper includes  data
collected by  the TESS  mission funded by  the NASA  Explorer Program.
This work  has made use of  data from the European  Space Agency (ESA)
mission {\it  Gaia} (\url{https://www.cosmos.esa.int/gaia}), processed
by  the {\it  Gaia}  Data Processing  and  Analysis Consortium  (DPAC,
\url{https://www.cosmos.esa.int/web/gaia/dpac/consortium}).     Funding
for the DPAC has been provided by national institutions, in particular
the  institutions   participating  in  the   {\it  Gaia}  Multilateral
Agreement. This research has made use of the services of the ESO
Science Archive Facility.

\end{acknowledgments} 

\facility{CTIO:1.5m, SOAR, Gaia, TESS}

\end{document}